\documentclass[aps,prl,twocolumn,showpacs,superscriptaddress,groupedaddress]{revtex4}  
\usepackage{graphicx}  
\usepackage{dcolumn}   
\usepackage{bm}        
\usepackage{amssymb}   
\usepackage{subfigure}
\usepackage{color,epsfig,rotate}
\usepackage{longtable}
\begin{document}



\title{Fermi-surface topology of the iron pnictide LaFe$_2$P$_2$}
\author{S. Blackburn}
\affiliation{D\'epartment de Physique and RQMP, Universit\'e de Montr\'eal,
Montr\'eal H3C 3J7, Canada}
\author{B. Pr\'evost}
\affiliation{D\'epartment de Physique and RQMP, Universit\'e de Montr\'eal,
Montr\'eal H3C 3J7, Canada}
\author{M. Bartkowiak}
\altaffiliation[Current address: ]{Laboratory for Developments and Methods, Paul Scherrer Institute, CH-5232 Villigen, Switzerland}
\affiliation{Hochfeld-Magnetlabor Dresden (HLD), Helmholtz-Zentrum
Dresden-Rossendorf and TU Dresden, D-01314 Dresden, Germany}
\author{O. Ignatchik}
\affiliation{Hochfeld-Magnetlabor Dresden (HLD), Helmholtz-Zentrum
Dresden-Rossendorf and TU Dresden, D-01314 Dresden, Germany}
\author{A.~Polyakov}
\affiliation{Hochfeld-Magnetlabor Dresden (HLD), Helmholtz-Zentrum
Dresden-Rossendorf and TU Dresden, D-01314 Dresden, Germany}
\author{T.~F\"orster}
\affiliation{Hochfeld-Magnetlabor Dresden (HLD), Helmholtz-Zentrum
Dresden-Rossendorf and TU Dresden, D-01314 Dresden, Germany}
\author{M. C\^ot\'e}
\affiliation{D\'epartment de Physique and RQMP, Universit\'e de Montr\'eal,
Montr\'eal H3C 3J7, Canada}
\author{G.~Seyfarth}
\affiliation{D\'epartment de Physique and RQMP, Universit\'e de Montr\'eal,
Montr\'eal H3C 3J7, Canada}
\affiliation{Laboratoire National des Champs Magn\'etiques Intenses (LNCMI), CNRS,
UJF, 38042 Grenoble, France}
\affiliation{Department of Physics and Astronomy, University of California Irvine,
Irvine, California 92697, USA}
\author{C. Capan}
\affiliation{Department of Physics and Astronomy, University of California Irvine,
Irvine, California 92697, USA}
\author{Z. Fisk}
\affiliation{Department of Physics and Astronomy, University of California Irvine,
Irvine, California 92697, USA}
\author{R. G. Goodrich}
\affiliation{Department of Physics, George Washington University,
Washington, DC, 20052, USA}
\author{I. Sheikin}
\affiliation{Laboratoire National des Champs Magn\'etiques Intenses (LNCMI), CNRS,
UJF, 38042 Grenoble, France}
\author{H. Rosner}
\affiliation{Max Planck Institute for Chemical Physics of Solids,
01187 Dresden, Germany}
\author{A. D. Bianchi}
\affiliation{D\'epartment de Physique and RQMP, Universit\'e de Montr\'eal,
Montr\'eal H3C 3J7, Canada}
\author{J. Wosnitza}
\affiliation{Hochfeld-Magnetlabor Dresden (HLD), Helmholtz-Zentrum
Dresden-Rossendorf and TU Dresden, D-01314 Dresden, Germany}
\date{\today}

\begin{abstract}
We report on a comprehensive de Haas--van Alphen (dHvA) study of the
iron pnictide LaFe$_2$P$_2$. Our extensive density-functional
band-structure calculations can well explain the measured
angular-dependent dHvA frequencies. As salient feature, we observe
only one quasi-two-dimensional Fermi-surface sheet, i.e., a hole-like
Fermi-surface cylinder around $\Gamma$, essential for $s_\pm$ pairing,
is missing. In spite of considerable mass
enhancements due to many-body effects, LaFe$_2$P$_2$ shows no
superconductivity. This is likely caused by the absence
of any nesting between electron and hole bands.

\end{abstract}

\pacs{71.18.+y, 71.27.+a, 74.25.Jb}
\maketitle

The very unique Fermi-surface topology of many iron-pnictide
superconductors stimulated a number of theories on the nature
of the pairing interactions in these materials. The degree of
nesting between quasi-two-dimensional hole and electron bands is
especially regarded as a key ingredient for a possible extended
$s$-wave pairing mechanism \cite{Mazin2008,Kor08,Cve09,maz10}.
Besides such intriguing topological nesting aspects of the
Fermi surface, other scenarios start from a more localized picture
to explain e.g.\ antiferromagnetism \cite{si08}. A dual description
of both itinerant and localized degrees of freedom of the Fe $d$
electrons is also considered \cite{dai12,Gorkov2013}.
In the latter cases, strong electronic correlations are expected
that lead to concomitant renormalized effective masses.

The precise knowledge of the electronic structure and many-body
interactions, therefore, is a prerequisite for testing the above
scenarios. The Fermi surface and band-resolved mass enhancement of
metals can ideally be investigated by quantum-oscillation studies
in combination with state-of-the-art band-structure calculations
\cite{Bergk08}. Indeed, such kind of investigations have been
reported for a number of iron pnictides, including several
members of the 122 family $X$Fe$_2$As$_2$ ($X$ = Ba, Sr, Ca)
\cite{seb08,ana09,har09} (see Refs.\ \onlinecite{Carrington2011, Coldea13}
for recent reviews). Important results on the topology
of the Fermi surfaces and many-body mass enhancements have been
obtained already, but more information is needed for gaining a
coherent picture on the band-structure properties relevant for
superconductivity in these materials.

Here, we report on a detailed de Haas-van Alphen (dHvA) study
combined with band-structure calculations of the 122 iron phosphide
LaFe$_2$P$_2$. Replacing in CaFe$_2$As$_2$ the divalent Ca
by trivalent rare earths of similar size leads, in the case of
La-doped CaFe$_2$As$_2$, to superconducting transition temperatures
up to about 47 K and a concomitant re-arrangement of the electronic
structure which could be driven by the interlayer As-As $p$-orbital
separation \cite{Paglione2012}. LaFe$_2$P$_2$, however, is a
nonsuperconducting analog of the iron-arsenide
superconductors of the 122 family.
LaFe$_2$P$_2$ is a Pauli-paramagnetic metal with trivalent
non-magnetic rare-earth La atoms \cite{Mor88} showing no signs
of any magnetic or superconducting order down to 20 mK. The electronic
structure, though anisotropic, is rather three dimensional and
characterized by a strongly corrugated quasi-two-dimensional
electron Fermi-surface sheet, as well as a donut-shaped, and a
strongly branched Fermi surface. Band-structure calculations describe
well the experimentally determined dHvA frequencies.

High-quality single crystals of LaFe$_2$P$_2$ were grown from
excess Sn flux \cite{REEHUIS1990}. The sample used for our
experiments has dimensions of $0.15 \times 0.1 \times 0.05$~mm$^3$.
We measured dHvA oscillations by use of capacitive cantilever
torque magnetometers, that could be rotated {\it in situ} around one axis.
The torque, $\vec{\tau}=\vec{M}\times \vec{B}$, results from the
interaction between the non-uniform magnetization $\vec{M}$ of
the sample with the applied magnetic field $\vec{B}$. The
measurements were performed both at the Dresden High Magnetic
Field Laboratory (HLD) using a superconducting 20-T magnet and
at the Grenoble High Magnetic Field Laboratory (LNCMI-Grenoble)
using a resistive water-cooled magnet reaching 34 T. The cantilevers
were mounted in both places directly in the $^3$He/$^4$He mixtures
of top-loading dilution refrigerators placed inside the magnets.

Since the resulting Fermi-surface topology is very sensitive to
details of the chosen approximations \cite{rem_calc}, we used
two different methods to calculate the electronic band
structure. Indeed, as seen below, both methods give somewhat different
dHvA frequencies, although the main features of the Fermi-surface
topology are nicely reproduced in both calculations. Using these
two different methods allowed us to get a better insight in the
reliability and the error bars of the extracted extremal orbits.
First, we applied the full-potential local-orbital
({\sc fplo}) code \cite{koe99} on a $24\times 24\times 24$ $k$ mesh.
Exchange and correlation potentials were estimated using the local
density approximation (LDA) \cite{per92}. Very similar results
were obtained when using the generalized gradient approximation (GGA)
\cite{Perdew1996}. For these calculations, we used the structural data
of Ref.\ \onlinecite{REEHUIS1990}, but we allowed the P atoms to relax their
$z$ position \cite{rem_Ppos}. Since it is known that in LDA $4f$ states
tend to shift too close to the Fermi energy \cite{rem_4f}, in a next
step we forced the La $4f$ states away from the Fermi energy by adding
the Coulomb correlation energy, expressed by $U$ \cite{rem_U}. This
LDA+$U$ formalism led only to minor changes in the band structure.

\begin{figure}
	\centering
	\includegraphics[width=0.95\columnwidth]{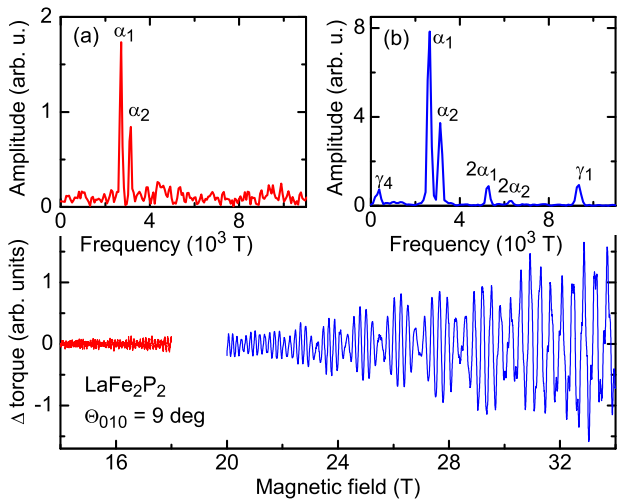}
	\caption{\label{dHvAexam} (Color online) The main panel shows the
background-subtracted torque signal obtained at about 30 mK in a
superconducting magnet from 14 to 18 T at the HLD and in a resistive
magnet between 20 and 34 T at the LNCMI-Grenoble. The fields
were aligned by about $9^\circ$ from the $c$ towards the $a$ axis.
The inset (a) shows the Fourier transformation of the HLD data,
the inset (b) that of the Grenoble data.}
\end{figure}

Further, the band structure was calculated through density
functional theory using the {\sc abinit} code \cite{Gonze2009}
with the projector-augmented wave formalism using the GGA
\cite{Perdew1996}.
The wave functions were calculated on a $16\times 16\times 16$
$k$-point grid in the first Brillouin zone using a plane-wave basis
up to an energy cutoff of 40~Ha (1088 eV). Here, the P atom
positions were relaxed as well. We used the experimentally
determined distance between the FeP planes in our calculations.

\begin{figure}
	\centering
	\subfigure{\label{fig:df_a}}
	  \includegraphics[width=0.7\columnwidth]{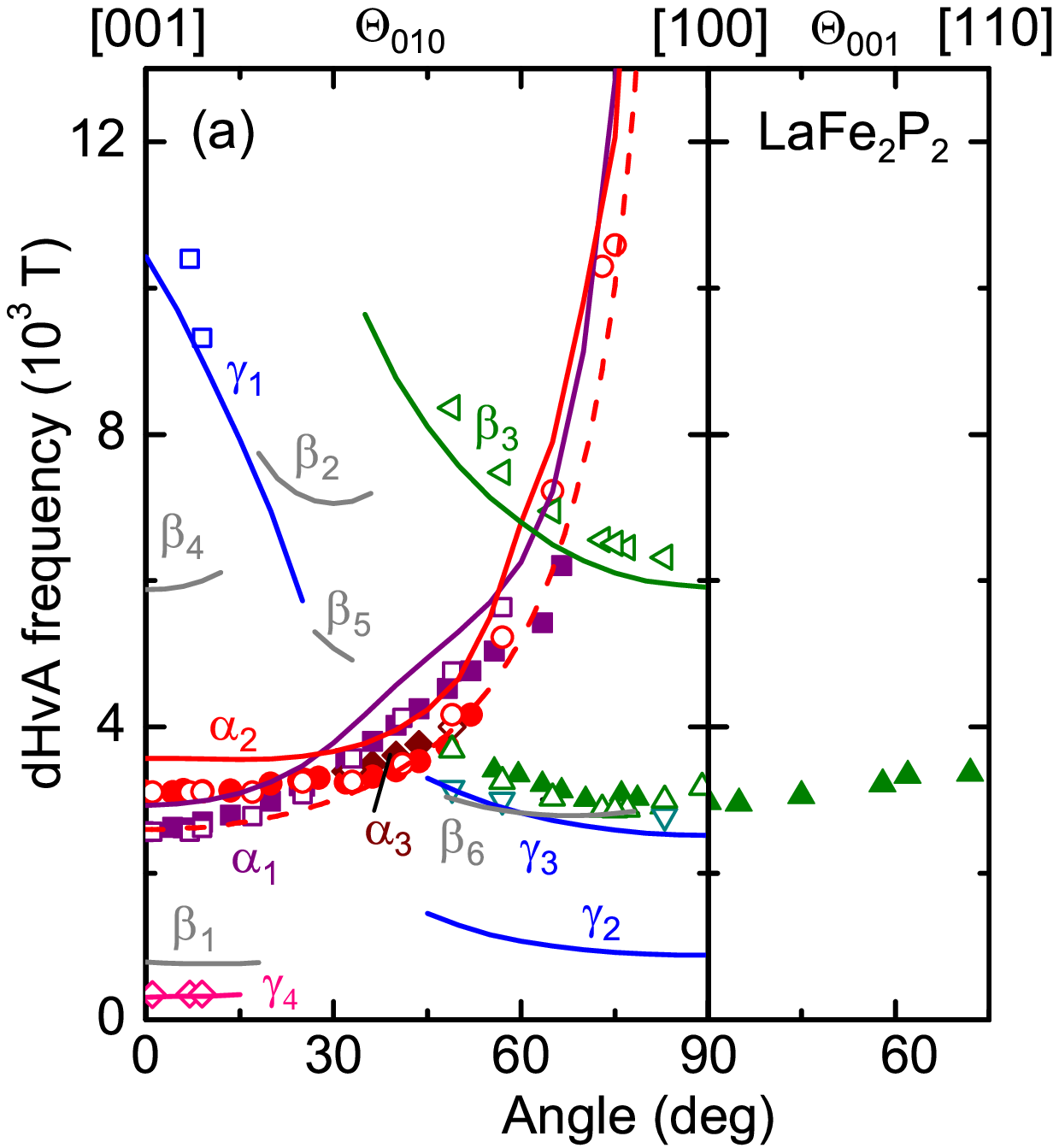}
	\subfigure{\label{fig:df_b}}
	  \includegraphics[width=0.6\columnwidth]{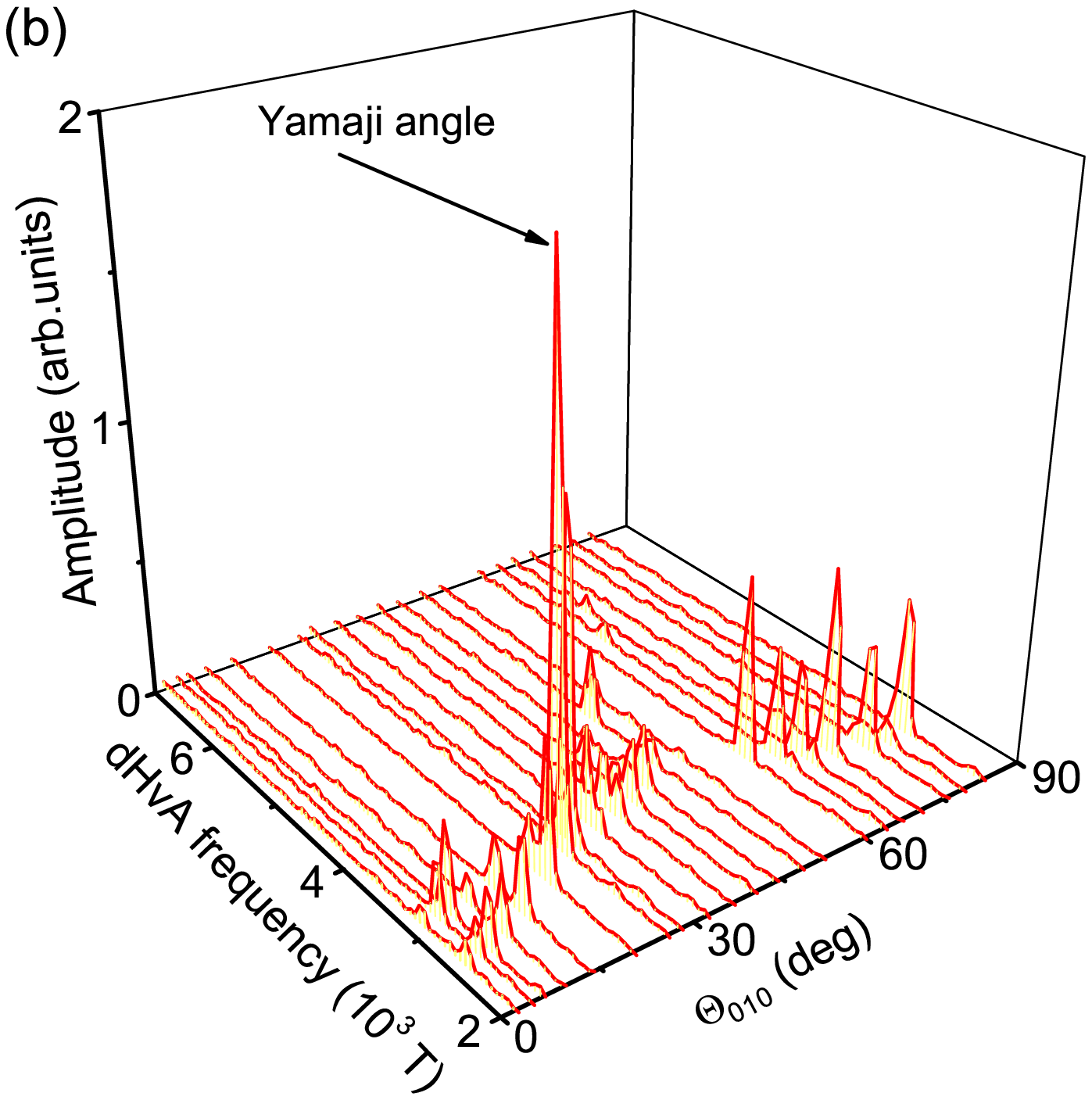}
	\caption{\label{fig:dhva_freq} (Color online) (a) Angular dependence
of the dHvA frequencies. Closed (open) symbols are experimental values
obtained at the HLD (LNCMI-Grenoble), while solid lines correspond to
the calculated extremal Fermi-surface orbits using the {\sc abinit}
code. The $\alpha$ orbits come from the electron-like
cylinder at $X$, the $\gamma$ orbits from the donut, and
the $\beta$ orbit from the multi-branched Fermi surface
(Fig.\ \ref{fig:FS_LDA}) (see main text for details).
The dashed curve shows a $1/\cos(\Theta)$ dependence expected for a
cylindrical Fermi surface. (b) 3D plot of the Fourier spectra obtained
for the field window between 14 and 18 T. The $\alpha_1$ and
$\alpha_2$ orbits merge at the Yamaji angle of about $25^\circ$
where the dHvA amplitude peaks clearly.}
\end{figure}

In our sample, we were able to resolve dHvA oscillations starting at
about 10~T on top of smoothly varying background torque signals. We
observed dHvA oscillations for all orientations of the applied
magnetic field. Seven independent dHvA frequencies could be resolved
over the whole angular range. In earlier measurements, only three
frequencies occurring over a restricted angular range were reported
for LaFe$_2$P$_2$ \cite{Mur09}. Typical examples of background-subtracted
(using fourth-order polynomials) dHvA torque signals are
shown in the main panel of Fig.\ \ref{dHvAexam}. For this
field orientation -- the magnetic field was rotated by about
$9^\circ$ from the $c$ towards the $a$ axis -- data up to 18 T
taken at the HLD in Dresden and from 20 up to 34 T obtained
in Grenoble are available \cite{rem_angle}. A beating
of the oscillating signal is evident which reflects the existence
of two slightly different dHvA frequencies. This is quantified
by the Fourier transformations of the signals. For the low-field
data [Fig.\ \ref{dHvAexam}(a)], only the two beating
frequencies at $F_{\alpha_1} = 2640(30)$~T and $F_{\alpha_2} =
3120(30)$~T could be resolved, while in the high-field data
[Fig.\ \ref{dHvAexam}(b)], two additional dHvA orbits appear, with
frequencies $F_{\gamma_4} \approx 370$~T and $F_{\gamma_1} = 9320(30)$~T,
besides the harmonics of $F_{\alpha_1}$ and $F_{\alpha_2}$.
The experimentally observed dHvA frequencies as a function
of the angles $\Theta_{010}$ (rotation from $c$ to $a$) and $\Theta_{001}$
(in-plane rotation) are shown in Fig.\ \ref{fig:df_a} together with data
obtained from the band structure calculated using the {\sc abinit}
code.

From the seven independent dHvA frequencies, some could only be
observed at the higher magnetic fields
available in Grenoble [open symbols in Fig.\ \ref{fig:df_a}].
Dominant are the dHvA signals labeled $\alpha_1$ and $\alpha_2$
(Fig.\ \ref{dHvAexam}) \cite{rem_alpha3}. The average of
these frequencies follows a $1/\cos(\Theta)$ dependence
[dashed line in Fig.\ \ref{fig:df_a}]
indicative of a corrugated quasi-two-dimensional Fermi surface.
Such a warped Fermi-surface cylinder leads to a peculiar well-defined
angular dependence of the dHvA frequencies and amplitudes. At the
so-called Yamaji angle \cite{Yamaji1989}, the point at which the two
dHvA frequencies merge, at about 25$^{\circ}$, a strong enhancement of
the dHvA amplitude is expected, since effectively
the extremal Fermi-surface area along the applied-field direction
stays constant -- as for an ideally two-dimensional metal -- meaning
that all electrons on that Fermi-surface sheet contribute to the dHvA
signal. This, indeed, fits nicely with the experimental observation as
shown in Fig.\ \ref{fig:df_b}.

\begin{figure}
	\centering
	\includegraphics[width=0.99\columnwidth]{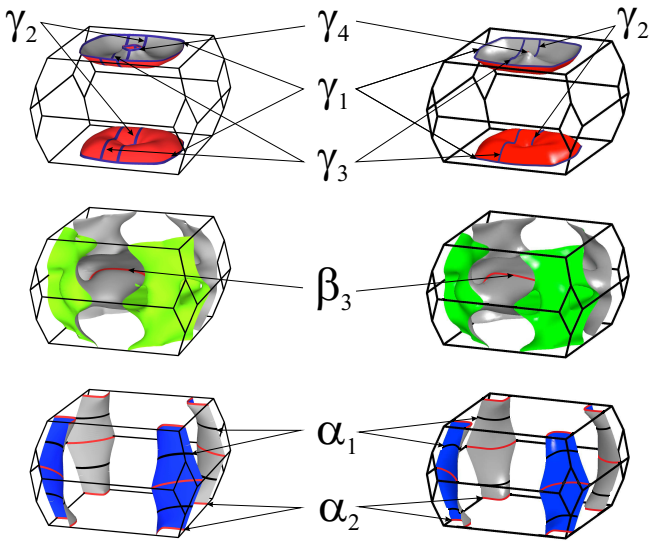}
	\caption{\label{fig:FS_LDA} (Color online) Fermi surface of LaFe$_2$P$_2$
calculated using the {\sc abinit} code (left) and the {\sc fplo} code
applying the LDA+$U$ approach (right). The upper Fermi-surface sheet is
hole-like, the other two electron-like. Solid lines mark
extremal orbits.}
\end{figure}

In order to assign the observed dHvA frequencies to extremal
Fermi-surface orbits, we performed comprehensive band-structure
calculations, as described above. Using the {\sc fplo} code (applying
the LDA+$U$ approach to the La $4f$ states) the resulting Fermi surfaces
together with the most prominent extremal orbits for fields along
[001] are shown in the right part of Fig.\ \ref{fig:FS_LDA}.
We obtain a very similar Fermi-surface topology when using the
{\sc abinit} code, shown in the left part of Fig.\ \ref{fig:FS_LDA}.
Our calculated Fermi surface more or less agrees with the
recently published one in Ref.\ \onlinecite{Moll2011}, but shows some
significant differences to the one reported in Ref.\ \onlinecite{Mur09}.

From the angular dependence of the extremal orbits the dHvA frequencies
were calculated. The result using the {\sc abinit} code is shown by
solid lines in Fig.\ \ref{fig:dhva_freq}(a). As a robust feature in
all our band-structure calculations, we find a strongly corrugated
cylinder at the Brillouin-zone corner, i.e., at the $X$ point (lower row
in Fig.\ \ref{fig:FS_LDA}). Such a cylinder is well known for the iron pnictides.
There is, however, no hole-like cylinder at the zone center $\Gamma$.
Therefore, the usually realized, at least partial, nesting between a cylinder
at $\Gamma$ and one at $X$ is absent -- and the electron pairing, essential
to the $s_\pm$ model \cite{Mazin2008,Kor08,Cve09,maz10}, is not possible.
Instead of a cylinder, a donut-shaped hole-like Fermi surface appears near $Z$
and a complicated ``monster'' fills more or less the whole Brillouin
zone.

\begin{figure}
	\begin{center}
	\includegraphics[width=0.98\columnwidth]{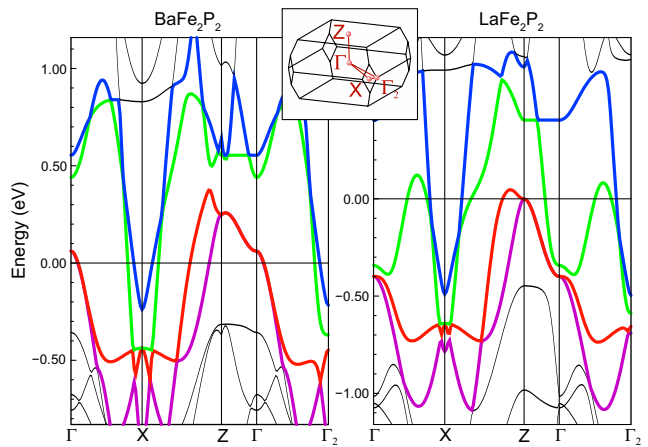}
	\end{center}
	\caption{\label{fig:BS_LDA_U} (Color online) Band structure of
BaFe$_2$P$_2$ (left) and LaFe$_2$P$_2$ (right) calculated using the
{\sc fplo} code. The Fermi energy
in the right panel is shifted by about 0.3 eV visualizing the
additional band filling caused by the trivalent La ions. The inset
shows the Brillouin zone with symmetry points.}
\end{figure}

Although mainly minor, some differences in the Fermi-surface topologies
are found when using the {\sc abinit} or {\sc fplo} code.
With the former, the hole of the donut is somewhat
larger, fitting perfectly to the experimental data [$\gamma_4$ orbit
in Fig.\ \ref{fig:dhva_freq}(a)]. On the other hand, the $\alpha$ orbits
of the cylinder at $X$ calculated using {\sc fplo} are about 7\,\%
smaller than in the {\sc abinit} code and fit the
data better. It is worth noting that the area of the cylinder depends
strongly on the distance between the FeP planes which would explain
the slightly enhanced theoretical dHvA frequencies. Anyway,
the agreement between experiment and our band-structure calculations
is very good [Fig.\ \ref{fig:dhva_freq}(a)]. For the cylinder at $X$,
the calculations predict crossings of the $\alpha_1$ and $\alpha_2$
orbits exactly at the experimentally observed angles. Close to [001],
we observe dHvA signals near 10~kT that can be assigned to the $\gamma_1$
orbit of the donut. The dHvA frequencies which are appearing at about
3000~T for a $\Theta_{010}$ above $50^\circ$ can be assigned
as well to the $\gamma_3$ donut as to an orbit of the
multi-branched $\beta$ surface (this $\beta$ orbit is not visible
in the visualization shown in Fig.\ \ref{fig:FS_LDA}).
For $\Theta_{010}$ between about $49^\circ$ and $83^\circ$
(Fig.\ \ref{fig:dhva_freq}), we could resolve the $\beta_3$ orbit
of the multi-branched Fermi surface which is nicely in line with the
calculations. Some of the predicted $\beta$ orbits appearing over
narrow angular ranges [gray lines in Fig.\ \ref{fig:dhva_freq}(a)]
have not been observed experimentally.

For comparison, we also performed band-structure calculations for
LaFe$_2$As$_2$ and BaFe$_2$P$_2$. In the first case, the Fermi surface
is almost unchanged which indicates that the $\Gamma$ hole-like cylinder
is destroyed by replacing Ba with La and not by the substitution of As
by P. Our results for BaFe$_2$P$_2$ are in agreement with those found
in the literature \cite{Shishido2010} with a hole-like cylinder in
the zone center. In Fig.\ \ref{fig:BS_LDA_U}, we show the band-structure
dispersions for BaFe$_2$P$_2$ (left) and LaFe$_2$P$_2$ (right). The
bands crossing the Fermi energy are highlighted by bold colored lines
corresponding to the coloring of the Fermi surfaces in Fig.\ \ref{fig:FS_LDA}.
The overall dispersions of these bands is similar for both
compounds. Although there are some clear differences in the dispersions
of the bands, the major effect is an upward shift of the Fermi energy
in LaFe$_2$P$_2$. This can be easily explained by the presence of
the trivalent La adding an additional electron to the Fermi sea, filling up
the bands to a higher level. As a second effect for a change of the
Fermi surface, we observe a surprisingly strong hybridization of the
La $5d$ electrons with the Fe $3d$ and the P $3p$ states. In contrast,
Ba states essentially do not contribute to the states near the Fermi
level in BaFe$_2$P$_2$. In consequence, the hole-like cylinders
around $\Gamma$ in BaFe$_2$P$_2$ originating in the red and magenta
bands disappear in LaFe$_2$P$_2$ (Fig.\ \ref{fig:BS_LDA_U}).
In addition, only one electron-like cylinder around $X$ remains for
LaFe$_2$P$_2$.

From the temperature dependence of the dHvA oscillation amplitudes we
determined the effective masses, as summarized in Table \ref{tab:mass}.
We find substantially smaller masses for the $\alpha$ orbits than
reported previously \cite{Mur09}. Anyhow, when comparing with the
calculated masses (using the {\sc fplo} code) \cite{rem_masses}, our
experimental results show that there exist considerable mass
renormalizations for all bands. These enhanced values are in agreement
with the strong correlations expected in the model where a Kondo-like
exchange mechanism is used to describe the dual nature of the iron $d$
electrons, being at the same time localized and itinerant
\cite{dai12,Gorkov2013}.

\renewcommand{\arraystretch}{1.2}
\begin{table}
\caption{\label{tab:mass} Comparison between the calculated (using
  the {\sc fplo} code) and measured dHvA frequencies and effective
  masses for different bands and field orientations in LaFe$_2$P$_2$.}
\begin{ruledtabular}
\begin{tabular}{lccccc}
 $\Theta_{010}$ & orbit & \multicolumn{2}{c}{measured}
  & \multicolumn{2}{c}{calculated}\\
\cline{3-4} \cline{5-6}
       &                   &  $F( \mathrm{T})$ & $m/m_{\mathrm{e}}$ &
       $F(\mathrm{T})$  &  $m_b/m_{\mathrm{e}}$  \\
\hline
  $0^\circ$   & $\alpha_1$ & 2570 & 1.5(1)  & 2730 & 0.68 \\
  $0^\circ$   & $\alpha_2$ & 3110 & 1.7(1)  & 3350 & 0.70 \\
  $75^\circ$  & $\gamma_3$ & 2880 & 3.1(1)  & 3100 & 0.98 \\
  $75^\circ$  & $\beta_3$  & 6510 & 3.6(2)  & 6110 & 1.53 \\
\end{tabular}
\end{ruledtabular}
\end{table}
\renewcommand{\arraystretch}{1}

When extracting the mass-enhancement factors, $\lambda =
m/m_b -1$, from the calculated, $m_b$, and measured, $m$, masses,
we obtain values for $\lambda$ between 1.2 and 2.2 for the
different bands. For other 122 materials such strong many-body
correlations are sufficient for superconductivity to occur
\cite{Carrington2011}. This strongly suggests that the absence
of the hole-like cylinder around $\Gamma$ indeed prevents the
material from becoming a superconductor. Our results indicate
that both sufficient nesting and strong many-body interactions
are needed for the appearance of superconductivity, most probably
of the extended $s$-wave type. However, more detailed models are
needed to elucidate which mass renormalization correlates with
the pairing potential.

In conclusion, our dHvA data combined with band-structure
calculations allowed to reveal a detailed picture of the
Fermi-surface topology of LaFe$_2$P$_2$. Compared to other 122
pnictides, the electron-like Fermi-surface cylinder in the
Brillouin-zone corner persists, whereas the hole-like cylinder
is absent. Instead a donut-like and a multi-branched Fermi-surface
sheet emerge. In spite of strong mass renormalizations, ascribed
to Kondo-like electronic correlations, no superconductivity
is found which can be explained by the absence of sufficient nesting
between the electron and hole parts of the Fermi surface.

The research at UdeM received support from the Natural Sciences and
Engineering Research Council of Canada (Canada), Fonds Qu\'eb\'ecois
de la Recherche sur la Nature et les Technologies (Qu\'ebec),
and the Canada Research Chair Foundation. Computational resources
were provided by Calcul Qu\'ebec and Compute Canada. Z.F.\ acknowledges
Grant No.\ NSF-DMR-0503361. Part of this work was supported by
EuroMagNET II (EU contract No.\ 228043) and by the
Deutsche Forschungsgemeinschaft (SPP 1458).


\begin{thebibliography}{99}

\bibitem{Mazin2008}
I. I. Mazin, D. J. Singh, M. D. Johannes, and M. H.
Du, Phys. Rev. Lett. {\bf 101}, 057003 (2008).

\bibitem{Kor08}
M. M. Korshunov and I. Eremin, Phys. Rev. B {\bf 78}, 140509(R) (2008).

\bibitem{Cve09}
V. Cvetkovic and Z. Tesanovic, EPL {\bf 85}, 37002 (2009).

\bibitem{maz10}
I. I. Mazin, Nature (London) {\bf 464}, 183 (2010).

\bibitem{si08}
Q. Si and E. Abrahams, Phys. Rev. Lett. {\bf 101}, 076401 (2008).

\bibitem{dai12}
P. Dai, J. Hu, and E. Dagotto, Nat. Phys. {\bf 8}, 709 (2012).

\bibitem{Gorkov2013}
L. P. Gor'kov, and G. B. Teitel'baum, Phys. Rev. B {\bf 87}, 024504 (2013).

\bibitem{Bergk08}
B. Bergk, V. Petzold, H. Rosner, S.-L. Drechsler, M. Bartkowiak,
O. Ignatchik, A. D. Bianchi, I. Sheikin, P. C. Canfield, and J. Wosnitza,
Phys. Rev. Lett. {\bf 100}, 257004 (2008).

\bibitem{seb08}
S. E. Sebastian, J. Gillett, N. Harrison, P. H. C. Lau, D. J.
Singh, C. H. Mielke, and G. G. Lonzarich, J. Phys.: Condens.
Matter {\bf 20}, 422203 (2008).

\bibitem{ana09}
J. G. Analytis, R. D. McDonald, J.-H. Chu, S. C. Riggs,
A. F. Bangura, C. Kucharczyk, M. Johannes, and I. R. Fisher,
Phys. Rev. B {\bf 80}, 064507 (2009).

\bibitem{har09}
N. Harrison, R. D. McDonald, C. H. Mielke, E. D. Bauer,
F. Ronning, and J. D. Thompson, J. Phys.: Condens. Matter
{\bf 21}, 322202 (2009).

\bibitem{Carrington2011}
A. Carrington, Rep. Prog. Phys. {\bf 74}, 124507 (2011).

\bibitem{Coldea13}
A. I. Coldea, D. Braithwaite, and A. Carrington, C. R. Physique
{\bf 14}, 94 (2013).

\bibitem{Paglione2012}
S. R. Saha, N. P. Butch, T. Drye, J. Magill, S. Ziemak, K. Kirshenbaum,
P. Y. Zavalij, J. W. Lynn, and J. Paglione,
Phys. Rev. B  {\bf 85}, 024525 (2012).

\bibitem{Mor88}
E. M\"orsen, B. D. Mosel, W. M\"uller-Warmuth, M. Reehuis,
and W. Jeitschko, J. Phys. Chem. Solids {\bf 49}, 785 (1988).

\bibitem{REEHUIS1990} M. Reehuis and W. Jeitschko, J. Phys.
Chem. Solids {\bf 51}, 961 (1990).

\bibitem{rem_calc}
For example, the appearance of the hole in the donut-shaped
Fermi surface depends on the exact positions of the P atoms.

\bibitem{koe99}
K. Koepernik and H. Eschrig, Phys. Rev. B {\bf 59}, 1743 (1999).

\bibitem{per92}
J. P. Perdew and Y. Wang, Phys. Rev. B {\bf 45}, 13244 (1992).

\bibitem{Perdew1996}
J.~P. Perdew, K. Burke, and M. Ernzerhof, Phys. Rev.
Lett. {\bf 77}, 3865 (1996).

\bibitem{rem_Ppos}
The P atom position was relaxed from $z = 0.3555$, the experimental
value of the $z$ position, to $z = 0.3477$.

\bibitem{rem_4f}
In consequence of this unphysical energy position of the $4f$
states, the Fermi surface may easily be distorted due to their
hybridization with the conduction electrons.

\bibitem{rem_U}
In order to safely shift the La $4f$ states away from the Fermi
energy we arbitrarily chose $U = 8$ eV.

\bibitem{Gonze2009}
X. Gonze, B. Amadon, P.-M. Anglade, J.-M. Beuken,
F. Bottin, P. Boulanger, F. Bruneval, D. Caliste, R. Caracas,
M. C\^ot\'{e}, et al., Comput. Phys. Commun.
{\bf 180}, 2582 (2009).

\bibitem{Mur09}
H. Muranaka, Y. Doi, K. Katayama, H. Sugawara,
R. Settai, F. Honda, T. D. Matsuda, Y. Haga, H. Yamagami, and Y. Onuki,
J. Phys. Soc. Jpn. {\bf 78}, 053705 (2009).

\bibitem{rem_angle}
We estimate the uncertainty in the correct sample alignment to be
about $2^\circ$.

\bibitem{rem_alpha3}
For some angles a third peak in the Fourier transformation appeared
next to the other $\alpha$ frequencies that most probably originates in
another extremal area of the strongly corrugated Fermi-surface cylinder.

\bibitem{Yamaji1989}
K. Yamaji, J. Phys. Soc. Jpn. {\bf 58}, 1520 (1989).

\bibitem{Moll2011}
P. J. W. Moll, J. Kanter, R. D. McDonald, F. Balakirev, P. Blaha,
K. Schwarz, Z. Bukowski, N. D. Zhigadlo, S. Katrych, K. Mattenberger,
J. Karpinski, and B. Batlogg,
Phys. Rev. B {\bf 84}, 224507 (2011).

\bibitem{Shishido2010}
H. Shishido, A. F. Bangura, A. I. Coldea, S. Tonegawa,
K. Hashimoto, S. Kasahara, P. M. C. Rourke, H. Ikeda,
T. Terashima, R. Settai, et al., Phys. Rev. Lett.
{\bf 104}, 057008 (2010).

\bibitem{rem_masses}
The calculated masses are very similar when using either the
{\sc fplo} or {\sc abinit} code.

\end{thebibliography}
\end{document}